\documentclass[a4paper,twoside]{article}
\usepackage{amssymb}
\usepackage{amsmath}
\usepackage{amsfonts}
\usepackage{graphicx}

\begin{document}

\title{\Large\bf{Loop quantum black hole}}

\author{\\ Leonardo
 Modesto
 \\[1mm]
\small{ Department of Physics, Bologna University V. Irnerio 46, I-40126 Bologna \& INFN Bologna, EU}\\
  \small{\&}\\
   \em\small{Centre de Physique Th\'eorique de Luminy,}\\ 
 \small{Universit\'e de la M\'editerran\'ee, F-13288 Marseille, EU}
   }

\date{\ } 
\maketitle

\begin{abstract}
In this paper we consider the Kantowski-Sachs space-time in Ashtekar variables
and the quantization of this space-time starting from the complete loop quantum gravity theory.
The Kanthowski-Sachs space-time coincides with the Schwarzschild black hole solution inside 
the horizon. By studying this model we can obtain information about the black hole
singularity and about the dynamics across the point $r=0$.
We studied this space-time in ADM variables in two previous papers where we showed
that the classical black hole singularity disappears in quantum theory.
In this work we study the same model in Ashtekar variables and we obtain  
a regular space-time inside the horizon region and that the dynamics 
can be extend further the classical singularity.  
\end{abstract}

\section*{Introduction}

In this work we study the space-time inside the horizon of a Schwazschild black hole
using the connection variables introduced by Ashtekar \cite{variables}.
We start the quantization program from the complete theory  of ``loop quantum gravity"
\cite{book} and we reduce the theory to consider an homogeneous but anisotropic 
minisuperspace model.

We studied the singularity problem in two previous papers \cite{work1} and  \cite{work2}.
There we used the same non Schr$\ddot{\mbox{o}}$dinger quantization procedure as
in the work of  V. Husain and O. Winkler on quantum cosmology. This formalism was
 introduced by Halvorson \cite{Fonte.Math} and also by A. Ashtekar, S. Fairhust and J. Willis \cite{AFW}. 
In the previous papers we used ADM formulation of general relativity and we obtained
that the quantum minisuperspace model for the black hole is singularity free; in fact in the first 
very simple model \cite{work1} we showed that the operator $1/r$ and so the curvature invariant 
$\mathcal{R}_{\mu \nu \rho \sigma} \, \mathcal{R}^{\mu \nu \rho \sigma} = 48 M^2 G_N^2/r^6$
are not divergent in the quantum theory. In the second paper \cite{work2} we considered   
a general two dimensional minisuperspace with space section of topology $\mathbf{R} \times \mathbf{S}^2$ (the Kantowski-Sachs space-time \cite{KS}) and we obtained that the  
inverse volume operator and the curvature invariant are singularity free in $r=0$ in quantum gravity.
In both models we have that at the quantum level the Hamiltonian constraint acts like a difference operator. A quantum extension on the other side of the classical singularity is straightforward.

We acknowledge Ioannis Raptis' s work on a mathematical approach to the singularity 
problem in general relativity \cite{I.R}.

The paper is organized as follows :
 in the first section we introduce the classical theory,
in particular we recall the invariant connection 1-form for the Kantowski-Sachs sapec-time.
We introduce the holonomies and the form of the Hamiltonian constraint and
the inverse volume operator in terms of such holonomies. We also express the quantity 
$1/r$ in terms of holonomies. This quantity is interesting because it is connected with 
the Schwarzschild curvature invariant 
$\mathcal{R}_{\mu \nu \rho \sigma} \, \mathcal{R}^{\mu \nu \rho \sigma}$
in the metric formulation of general relativity.
In the second section we quantize the system. In particular we
calculate explicitly the spectrum of the inverse volume operator, the spectrum 
of the operator $1/r$ and the solution of the Hamiltonian constraint in the 
dual to the kinematical Hilbert space. 

\section{Classical theory}
In this section first of all we summarize the fundamental Hamiltonian formulation of
``loop quantum gravity" 
\cite{book} then we recall the formulation of the Kantowski-Sachs space-time
in Ashtekar variables \cite{Bombelli}.   
At this point we define the holonomies for 
the connection and we rewrite the classical Hamiltonian constraint and 
the inverse volume operator in terms of holonomies.

\subsection{Loop quantum gravity preliminary}
The classical starting point of loop quantum gravity (LQG) is the Hamiltonian formulation of 
general relativity. Historically in the ADM Hamiltonian 
formulation of the Einstein theory the fundamental variables are the three-metric 
$q_{ab}$ of the spatial section $\Sigma$ of a foliation of the $4-d$ manifold 
$\mathcal{M} \cong \mathbb{R} \times \Sigma$ and the extrinsic curvature $K_{ab}$.
In LQG the fundamental variables are the Ashtekar variables: they consist  
of an $SU(2)$ connection $A_a^i$ and the electric field $E_i^a$, where $a, b, c, \dots = 1, 2, 3$ are tensorial indexes on the spatial section and $i, j, k, \dots = 1, 2, 3$ are indexes in the 
$su(2)$ algebra.   
The density weighted triad $E^a_i$ is define in terms of the triad $e^i_a$ by the 
relation $E^a_i = \frac{1}{2} \epsilon^{abc} \, \epsilon_{ijk} \, e^j_b \, e^k_c$. And the 
metric can be expressed in terms of the triad by $q_{a b} = e^i_a \, e^j_b \, \delta_{ij}$.

We can express also the metric in terms of the weighted triad as 
\begin{equation}
q \, q^{a b} = E^a_i \, E^b_j \, \delta^{ij}, \,\,\,\, q = \sqrt{\mbox{det}(q_{a b})}.
\end{equation}
The relation between the variables $(A^i_a, E^a_i)$ and the ADM variables 
$(q_{ab}, K_{ab})$ is 
\begin{eqnarray}
A^i_a = \Gamma^i_a + \gamma \, K_{ab} E^b_j \delta^{i j}
\end{eqnarray} 
where $\gamma$ is the Immirzi parameter and $\Gamma^i_a$ is the spin 
connection, being the solution of Cartan's equation: 
$\partial_{[a} e^i_{b]} + \epsilon^i_{j k} \, \Gamma^j_{[a} e^k_{b]} = 0$.

The action in the new variables is 
\begin{eqnarray}
S = \frac{1}{k} \int dt \int_{\Sigma} d^3 x \left[- 2 T_r(E^a A_a) - N H - N^a H_a - N^i G_i \right],
\label{action}
\end{eqnarray}  
where $N^a$ is the shift vector, $N$ is the lapse function and $N^i$ is the Lagrange
multiplying for the Gauss constraint $G_i$. The functions $H$, $H_a$ and $G_i$
are respectively the Hamiltonian, diffeomorphism and Gauss constraints . 

The constrains are given by 
\begin{eqnarray}
&& H(E^a_i, A^i_a) = - \frac{4}{\sqrt{|\mbox{det}(E)}|} \, Tr \left[F_{ab} \, E^a E^b \right] 
- 2(1+ \gamma^2) \frac{E^a_i E^b_j K^i_{[a} K^j_{b]}}{\sqrt{|\mbox{det} (E)}|} \nonumber \\
&& H_b(E^a_i, A^i_a) = E^a_j \, F^j_{ab} - (1+ \gamma^2) K^i_a G_i \nonumber \\
&& G_i(E^a_i, A^i_a) = \partial_a E^a_i + \epsilon_{ij}^k \, A^j_a E^a_k, 
\label{constraints}
\end{eqnarray}
and the curvature field straight is $F_{ab} = \partial_a A_b - \partial_b A_a + \left[A_a, A_b \right]$.

 After this brief introduction to "loop quantum gravity" we can study the 
 "Kantowski-Sachs" symmetric reduced model.
   
 \subsection{Kantowski-Sachs space-time in Ashtekar variables}
At this point after we recalled the basic ingredients of LQG we can get back to study
the Kantowski-Sachs space-time. This is a simplified model of a generical 
homogeneous but anisotropic space-time of coordinates $(t,r,\theta, \phi)$. 

In general a homogeneous but anisotropic space-time of spatial section $\Sigma$
of topology $\Sigma \cong \mathbb{R} \times S^2$ is characterized by a invariant 1-form 
connection $A_{[1]}$ of the form \cite{BojThiemann}
\begin{eqnarray}
A_{[1]} = A_r(t) \, \tau_3 \, dr + (A_1(t) \, \tau_1 + A_2(t) \tau_2) \, d \theta +
           (A_1(t) \, \tau_2 - A_2(t) \tau_1) \sin \theta \, d \phi + \tau_3 \, \cos \theta \, d \phi.
           \label{symconnection}
\end{eqnarray} 
The $\tau_i$ generators of the $SU(2)$ fundamental representation are relate to the
Pauli $\sigma_i$ matrix by $\tau_i = -  \frac{i}{2} \sigma_i$.

On the other side the dual invariant densitized triad is 
\begin{eqnarray}
E_{[1]} = E^r(t) \, \tau_3 \, \sin \theta \, \frac{\partial}{\partial r} + (E^1(t) \, \tau_1 + E^2(t) \, \tau_2)
\, \sin \theta \, \frac{\partial}{\partial \theta}  +
           (E^1(t) \, \tau_2 - E^2(t) \tau_1) \frac{\partial}{\partial \phi}.
           \label{symtriad}
\end{eqnarray} 

In this case the space-time is homogeneous and the diffeomorphism constraint is
automatically satisfied; instead the Gauss constraint and the Euclidean part of the 
Hamiltonian constrains are similar to 
\begin{eqnarray}
&& G \sim A_1 E^2 - A_2 E^1 \nonumber \\
&& H_E = \frac{2 \sin \theta \, \, \mbox{sgn}[\mbox{det}(E_{[1]})]}{\sqrt{|E^r|[(E^1)^2 + (E^2)^2]}} \, \Big[2 A_r E^r (A_1 E^1 + A_2 E^2) +
   \Big((A_1)^2 + (A_2)^2 - 1\Big)[(E^1)^2 + (E^2)^2 \Big] \nonumber \\
   &&
   \label{gauss}
\end{eqnarray}

In this paper we want to study the Kantowski-Sachs space-time 
with space section of topology $R \times S^2$; the connection $A_{[1]}$ 
is more simple than in (\ref{symconnection}) with $A_2 = A_1$, and in the triad 
(\ref{symtriad}) we can choose the gauge $E^2 = E^1$ \cite{Bombelli}. There is 
another residual gauge freedom on the pair $(A_1, E^1)$. This is a discrete transformation
$P: (A_1, E^1) \rightarrow (-A_1, -E^1)$; we have to fix this symmetry on the Hilbert space.

The Gauss constraint in 
(\ref{gauss}) is automatically satisfied and the Euclidean part of the Hamiltonian constraint becomes
\begin{eqnarray}
H_E = \frac{2 \sqrt{2} \sin \theta \,\,\mbox{sgn}(E^r)}{\sqrt{|E^r|}|E^1|} \, \Big[2 A_r E^r A_1 E^1 + ( 2 (A_1)^2  - 1)(E^1)^2 \Big]
 \label{hamiltonian}
\end{eqnarray}
from now on we redefine $E^r \equiv E$ and $A_r \equiv A$. 

In the metric formalism the kantowski-Sachs space-time assume the form
\begin{eqnarray}
ds^2 = - dt^2 + a^2 (t) dr^2 + b^2 (t)  (\sin^2 \theta d\phi^2 + d \theta^2),
\label{metricab}
\end{eqnarray}
where $a(t)$ and $b(t)$ are two independent functions of the time coordinate $"t"$.

The connection between the metric and density triad formulation is 
\begin{equation}
q_{a b} = \left(\begin{array}{ccc}
                          \frac{2 (E^1)^2}{|E|}     &     0                                   &    0  \\
                         0    &     |E|    &    0  \\
                         0    &     0                                   &   |E| \, \sin^2 \theta

\end{array}   \right).
\label{metric}
\end{equation}

Using (\ref{metric}) we can calculate the volume of the spatial section $\Sigma$
\begin{eqnarray}
V = \int dr \, d \phi \, d \theta \, \sqrt{q} = 4 \pi \sqrt{2} R \sqrt{|E|} |E^1|,
\label{Volume}
\end{eqnarray}
where $R$ is a cut-off on the space radial coordinate. We can use also radial densities 
because the model is homogeneous and all the following results remain identical. In another way, 
the spatial homogeneity enable us to fix a linear radial cell $\mathcal{L}_r$ and restrict all 
integrations to this cell \cite{Boj}.   

To obtain the classical symplectic structure of the phase space we can calculate the
first term of the action (\ref{action}) introducing in it our variables. 

This very simple calculation produces
\begin{eqnarray}
&& S = \frac{1}{\kappa} \int d t \int d r \, d \phi \, d \theta \, \sin \theta \,
\left[Tr \left( -2 E^a A_a \right) + \dots \right] = 
\frac{R}{\kappa} \int d t \int d \phi \, d \theta \, \sin \theta \, \left[E A + 4 E^1 A_1 + \dots \right].\nonumber \\
&&
\label{actionred}
\end{eqnarray}

Concerning what we said before we can restrict the linear radial cell to the Planck length
$l_P$  and so we can take $R \equiv l_P$. 
Observing the reduced action (\ref{actionred}) we obtain the simplectic structure
of the classical phase space. The phase space consists of two canonical 
pairs $A, E$ and $A_1, E^1$ and the simplectic structure is given by the
poisson brackets, $\{A, E\} = \frac{\kappa}{l_P}$ and $\{A_1, E^1\} = \frac{\kappa}{4 l_P}$.
The coordinates and the momentums have dimensions: $[A] = L^{-1}$, $[A_1] = M^0$,
$[E] = L^2$ and $[E^1] = L$.  
 
\subsection{Holonomies, curvature and Hamiltonian constraint} 
In this section we introduce the fundamental classical ingredients necessary 
to quantize the theory.
To define the holonomies we don't consider the field independent term 
in the connection; this means that we define the model on particular spin networks
which are based on graph made only of radial edges, 
or of edges along circles in the $\theta$-direction or at $\theta = \pi/2$. 
Considering this graph type we can omit the term ``$\tau_3 \, \cos \theta \, d \phi$" in the
holonomy in the ``$\phi$" direction. 

We introduce the background triad $^o e^a_I = \mbox{diag}(1, 1, \sin^{-1} \theta)$ and co-triad 
$^o \omega^I_a = \mbox{diag}(1, 1, \sin \theta)$, and we define the holonomy
  \begin{eqnarray}
h = \mbox{e}^{\int A_{[1]}} = \mbox{e}^{ \int A_a d x^a} = \mbox{e}^{\int A^i_a \, \frac{d x^a}{d \lambda} \, \tau_i d \lambda}
= \mbox{e}^{\int A^i_I  \,^o\omega^I_a \,^oe_J^a  \, u^J \, \tau_i d \lambda} = 
\mbox{e}^{\int A^i_I  \, u^I \, \tau_i d \lambda} , 
\label{holonomies}
\end{eqnarray}
where $u^a = \frac{d x^a}{d \lambda} = (R \, \frac{dr^{\prime}}{d \lambda}, \frac{d \theta}{d \lambda}, \frac{d \phi}{d \lambda})$ and $u^I =  ^o \omega^I_a \, u^a$, $r^{\prime}$ is a dimensionless radial coordinate and $R$ is the scale introduced in (\ref{Volume}). In the rest of the paper as we 
wrote in the previous section we will take $R=l_P$ to simplify the formulas. 

We can define now the holonomy along the direction  ``$I \, $" 
\begin{eqnarray}
&& h_I= \exp \int A^i_I  \, \tau_i d \lambda , \nonumber \\
&& h_1 = \exp \int A^i_1 \tau_i d \lambda = \exp[ A \mu_0 l_P \, \tau_3] , \nonumber \\
&& h_2 = \exp \int A^i_2 \tau_i d \lambda = \exp [A_1 \mu_0  \, (\tau_2 + \tau_1)] , \nonumber \\
&& h_3 = \exp \int A^i_3 \tau_i d \lambda = \exp [A_1 \mu_0   \, (\tau_2 - \tau_1) ], 
\label{holonomiyI}
\end{eqnarray}
and introducing the normalized vectors $n_1^i = (0, 0, 1)$, $n_2^i = \frac{1}{\sqrt{2}} (1, 1, 0)$,
  $n_3^i = \frac{1}{\sqrt{2}} (-1, 1, 0)$ we can rewrite $h_I$ as
  \begin{eqnarray}
&& h_I= \exp \left( \bar{A}_I \, n^i_I  \, \tau_i \right)= \cos\Big(\frac{\bar{A}_I}{2}\Big) + 2 n^i_I \, \tau_i \, \sin\Big(\frac{\bar{A_I}}{2}\Big)  , \nonumber \\
&& h_1 = \cos\Big(\frac{A \mu_0 l_P}{2}\Big) +  2 \tau_3 \, \sin\Big(\frac{A \mu_0 l_P}{2}\Big)  , \nonumber \\
&& h_2 = \cos\Big(\frac{\sqrt{2} A_1 \mu_0}{2}\Big) +\sqrt{2} (\tau_2 + \tau_1)\, \sin\Big(\frac{\sqrt{2} A_1 \mu_0}{2}\Big)  , \nonumber \\
&& h_3 =  \cos\Big(\frac{\sqrt{2} A_1 \mu_0}{2}\Big) +\sqrt{2} (\tau_2 - \tau_1) \, \sin\Big(\frac{\sqrt{2} A_1 \mu_0}{2}\Big)  
\label{holonomiyI2}
\end{eqnarray}

We remember that the Hamiltonian constraint can be written in terms of the curvature
$F_{ab}$ and the Poisson bracket between $A_a$ and the volume $V$. 
The Euclidean part of the Hamiltonian constraint becomes   
\begin{eqnarray}
H_E = - 2 \int d^3 x \, \epsilon^{a b c} \, \mbox{Tr} \Big[F_{a b} \, \{A_c, V\} \Big].
\label{hamiltonianE}
\end{eqnarray}

We now rewrite the hamiltonian constraint (\ref{hamiltonianE}) in terms of holonomies.
We start to define the curvature $F_{a b}$ using the holonomies (\ref{holonomiyI})
 \begin{eqnarray}
F_{a b}^i \, \tau_i = \,^o \omega^I_a \,^o \omega^J_b \, 
\left[\frac{h_I h_J h_I^{-1} h_J^{-1}h_{[IJ]} -1}{\epsilon(I) \epsilon(J)}\right],
\label{curvature}
\end{eqnarray}
where $\epsilon(1) = R \, \mu_0$, $\epsilon(2) = \epsilon(3) =  \mu_0$. 
We have introduced also the field independent holonomy $h_{[IJ]}$ to obtain 
the correct classical curvature in the limit $\epsilon(I) \rightarrow 0$. This term
is defined by 
\begin{eqnarray}
 h_{[IJ]} = \exp(- \mu_0^2 \, C_{I J} \, \tau_3) = \cos\left(\frac{\mu_0^2 \, C_{IJ}}{2}\right) - 2 \, \tau_3 \, \sin\left(\frac{\mu_0^2 \, C_{IJ}}{2}\right) ,
 \end{eqnarray}
where $C_{IJ}$ is a constant antisimmetric matrix, 
$C_{IJ} = \delta_{2 I} \delta_{3 J} - \delta_{3_I} \delta_{2 J}$.

The other term we need
is ``$\tau_i \, \{ A^i_a, V \}$" and in terms of holonomies becomes 
\begin{eqnarray}
\tau_i \, \{ A^i_a, V \} = \,^o \omega^I_a \, \frac{h_I^{-1} \{h_I, V\}}{\epsilon(I)}.
\label{AV}
\end{eqnarray}

At this point we have all the elements to define the Euclidean part of the Hamiltonian constraint
\begin{eqnarray}
 H_E & = & - 2 \int d^3 x \, \epsilon^{a b c} \, \,^o \omega^I_a \,^o \omega^J_b \,^o \omega^K_c \, \mbox{Tr} \left[\frac{\left(h_I h_J h_I^{-1} h_J^{-1} h_{[IJ]} - 1 \right) \, h_K^{-1} \{h_K, V \}}{\epsilon(I) \epsilon(J) \epsilon(K)}  \right] \nonumber \\
 & =  &  - 2 \int d^3 x \, \sqrt{\,^o q} \, \epsilon^{I J K} \, \mbox{Tr} \left[\frac{\left(h_I h_J h_I^{-1} h_J^{-1} h_{[IJ]} - 1 \right) \, h_K^{-1} \{h_K, V \}}{\epsilon(I) \epsilon(J) \epsilon(K)}  \right].
\label{hamiltonianEreg}
\end{eqnarray}
Recalling the $\epsilon(I)$'s definitions and that $\int d^3x \,\sqrt{\,^o q} = 4 \pi l_p$ we 
can simplify the expression (\ref{hamiltonianEreg})
\begin{eqnarray}
 H_E  =  - \frac{8 \pi}{\mu_0^3} \, \sum_{I J K}\, \epsilon^{I J K} \, \mbox{Tr} \left[h_I h_J h_I^{-1} h_J^{-1} h_{[IJ]} \, h_K^{-1} \{h_K, V \}  \right],
\label{hamiltonianEreg2}
\end{eqnarray}
where we have also eliminated the identity term
because it doesn't contribute when traced. 

\subsection{Inverse volume \& Schwarzschild curvature invariant \\in terms of holonomies}
We can define the inverse volume operator following Thiemann. We remember
that the volume operator is $V = 4 \pi \sqrt{2} l_P |E|^{1/2} |E^1|$ and we define 
$\bar{V} = \frac{\sqrt{\mbox{det}(q)}}{\sin \theta} =  |E|^{1/2} |E^1|$.

A useful property to define the inverse volume operator is 
 \begin{eqnarray}
  \frac{\mbox{det}(e^i_a)}{\,\,\,(\mbox{det}(q))^{\frac{3}{4}}} =
  \frac{\sqrt{\mbox{det}(q)}}{\,\,\,(\mbox{det}(q))^{\frac{3}{4}}} = \frac{1}{\sqrt{\bar{V}}},
 \end{eqnarray}
 using this relation and the identity 
  $e^i_a = - \frac{(16)^{\frac{1}{3}} l_P}{\kappa} \, \{A^i_a, V\}$ we obtain 
 \begin{eqnarray}
 \frac{\mbox{sgn}(E)}{\sqrt{\bar{V}}} = - \frac{64}{3} \, \left(\frac{l_P}{\kappa} \right)^3 \,
 \epsilon_{i j k} \, \epsilon^{a b c} \, 
 \{A^i_a, \bar{V}^{\frac{1}{2}} \}
  \{A^j_b, \bar{V}^{\frac{1}{2}} \}
   \{A^k_c, \bar{V}^{\frac{1}{2}} \}.
   \label{ClassicVol}
      \end{eqnarray} 
 To rewrite the volume in terms of holonomies we use extensively the holonomy 
 version of the triad 
 \begin{eqnarray}
 e^i_a  =  \frac{2 (16)^{\frac{1}{3}} l_P}{\kappa} \,^o \omega^I_a \, 
 \frac{\mbox{Tr}[\tau^i h_I^{-1} \{h_I, \bar{V} \}]}{\epsilon(I)}.
 \label{holtriad}
 \end{eqnarray}
 At this point using the definition (\ref{holtriad}) in (\ref{ClassicVol}) we obtain the final version
 of the inverse volume operator in terms of holonomies
  \begin{eqnarray}
&&\hspace{-0.7cm} \frac{\mbox{sgn}(E)}{\sqrt{\bar{V}}}  =  
\left(\frac{512 \, l_P^3}{3 \, \kappa^3} \right) \, 
 \epsilon_{i j k} \, \sum_{I J K} \frac{\epsilon^{I J K} \, \mbox{Tr}[\tau^i h_I^{-1} \{h_I, \bar{V}^{\frac{1}{2}} \}] \,  \mbox{Tr}[\tau^j h_J^{-1} \{h_J, \bar{V}^{\frac{1}{2}} \}] \,\mbox{Tr}[\tau^k h_K^{-1} \{h_K, \bar{V}^{\frac{1}{2}} \} ]}{\epsilon(I) \, \epsilon(J) \epsilon(K)}. \nonumber \\
 &&
    \label{ClassicVol2}
      \end{eqnarray}

 We consider now another operator that is useful to understand the black hole singularity. 
 In the metric formulation of the Schwarzschild black hole the invariant quantity 
 $\mathcal{R}_{\mu \nu \rho \sigma} \, \mathcal{R}^{\mu \nu \rho \sigma} \sim  1/r^6$ 
 diverges in $r=0$ where the singularity is localized.
 The relevant quantity $1/r^2 = 1/b^2(t)$ in metric formulation is given by 
 $1/|E|$ in Ashtekar variables.
 
 Now we express $1/\sqrt{|E|}$ in terms of holonomies 
 \begin{eqnarray}
 \frac{\mbox{sgn}(E) \, \mbox{sgn}(E^1)}{\sqrt{|E|}} =\frac{64 \, l_P}{\mu_0^2 \, \kappa^2}
 \, \mbox{Tr} \left[\tau_3 h_1^{-1} \{h_1, \bar{V}^{\frac{1}{2}} \} \right] \, 
 \mbox{Tr} \left[(\tau_2+\tau_1) h_2^{-1} \{h_2, \bar{V}^{\frac{1}{2}} \} \right]
  \label{CurvInv}
 \end{eqnarray}
We will use the quantities (\ref{ClassicVol2}) and (\ref{CurvInv}) in the next section where we study the black hole singularity for $E=0$ in the quantum theory. 
 
\section{Quantum Theory}
We want quantize the symmetry reduced theory using techniques from the
full theory (loop quantum gravity).   
The Hilbert space for the Kantowski-Sachs minisuperspace model in 
Ashtekar variables is the tensor product of two Hilbert spaces, one for 
the anisotropy operator $\hat{E}^1$ and the other for the operator $\hat{E}$ : 
$\mathcal{H}_E \otimes \mathcal{H}_{E^1}$; this is the kinematical Hilbert
space ($\mathcal{H}_{kin}$) for the reduced model obtained representing 
the almost periodic function algebra (holonomy algebra for the homogeneous space-time) 
using the Bohr compactification of the real line. 

A basis on this space is given by
\begin{eqnarray}
|\mu_E,  \mu_{E^1} \rangle \equiv |\mu_E \rangle \otimes |\mu_{E^1} \rangle 
\rightarrow \langle A|\mu_E\rangle \otimes \langle A_1|\mu_{E^1} \rangle =  e^{\frac{i \mu_E \, l_P \, A}{2}} \otimes e^{\frac{i \mu_{E^1} \, A_1}{\sqrt{2}}}.
\label{basis}
\end{eqnarray} 

The basis states (\ref{basis}) are normalizable in contrast to the standard quantum mechanical 
representation and they satisfy 
\begin{eqnarray}
\langle \mu_E, \mu_{E^1}| \nu_E, \nu_{E^1} \rangle = \delta_{\mu_E, \nu_E} \, \delta_{\mu_{E^1}, \nu_{E^1}}
\end{eqnarray}
We can quantize the theory using the standard quantization procedure 
$\{\, , \,\} \rightarrow - \frac{i}{\hbar} [\, , \, ]$. 
We recall the fundamental difference from the standard Schre$\ddot{\mbox{o}}$dinger 
quantization program.  
In the ordinary Schr$\ddot{\mbox{o}}$edinger 
representation of the Weyl-Heisenberg algebra the classical fields $A$ and $A_1$ 
translate in operators. In loop quantum gravity (or polymer representation 
of the Weyl algebra) on the contrary the operators $A$ and $A_1$ don't exist.  
We can't promote the Poisson brackets to commutators 
$[\hat{A}, \hat{E}] = i \sqrt{l_P}$, $[\hat{A}_1, \hat{E}^1] = i \sqrt{l_P}/4$, 
but instead the quantum fundamental operators are $\hat{E}, \hat{E}^1, \hat{h}_I$.

The momentum operators can be represented on the Hilbert space by 
\begin{eqnarray}
&& \hat{E} \rightarrow - i l_P \frac{d}{d A}, \nonumber \\
&& \hat{E}^1 \rightarrow - i \frac{l_P}{4} \frac{d}{d A_1},
\label{EE1}
\end{eqnarray}

It is very simply to calculate the spectrum of these two momentum operators on the Hilbert space 
basis, and so we obtain 
\begin{eqnarray}
&& \hat{E} |\mu_E, \mu_{E^1} \rangle = \frac{\mu_E \,  l_P^2}{2} |\mu_{E}, \mu_{E^1} \rangle,\nonumber \\
&& \hat{E}^1 |\mu_E, \mu_{E^1} \rangle = \frac{\mu_{E^1} \, l_P}{4 \sqrt{2}} |\mu_E, \mu_{E^1} \rangle.
\label{diagEE1}
\end{eqnarray}

 We have to fix the residual gauge freedom on the Hilbert space. We consider the
 operator $\hat{P} : |\mu_E, \mu_{E^1} \rangle \rightarrow |\mu_E, - \mu_{E^1} \rangle$
 and we impose that only the invariant states (under $\hat{P}$) are in the kinematical
 Hilbert space. The states in the Hilbert space are of the form 
 $\frac{1}{2} \left[|\mu_E, \mu_{E^1} \rangle + |\mu_E, - \mu_{E^1} \rangle\right]$.
 
 In the following, we calculate the action of the volume, of the inverse volume and of the 
Schwarzschild curvature invariant operators on the kinematical Hilbert space, 
obtaining a bounded spectrum for these operators.
  
\subsection{Inverse volume \& $1/\sqrt{|E|}$ operators}
In this section we  study the black 
hole singularity problem in loop quantum gravity . 
We recall that the particular form of the volume for the space-time 
inside the horizon is $V = 4 \sqrt{2} \, \pi l_P \, \sqrt{|E|} |E^1|$.

The action of the volume operator on the basis states is 
\begin{eqnarray}
\hat{V} | \mu_E, \mu_{E^1} \rangle = \frac{4 \pi \, R \, l_P^2}{\sqrt{2}} \,\sqrt{|\mu_E|} \, |\mu_{E^1}| |\mu_E, \mu_{E^1}\rangle.
\end{eqnarray}
Now we will show that the operator $1/\sqrt{\bar{V}}$ doesn't diverge in quantum theory.  

The operator version of the quantity $\mbox{sgn}(E)/\sqrt{\bar{V}}$ defined in the
formula (\ref{ClassicVol2}) is 
 \begin{eqnarray}
&& \widehat{\frac{\mbox{sgn}(E)}{\sqrt{\bar{V}}} } =  \frac{512 \, i}{3 \, l_P^4 \mu_0^3}  \, 
 \epsilon_{i j k}  \sum_{I J K} \epsilon^{I J K} \, \mbox{Tr}\left[\tau^i \hat{h}_I^{-1} [\hat{h}_I, \hat{\bar{V}}^{\frac{1}{2}}] \right] \,  \mbox{Tr}\left[\tau^j \hat{h}_J^{-1} [\hat{h}_J, \hat{\bar{V}}^{\frac{1}{2}}] \right] \,\mbox{Tr}\left[\tau^k \hat{h}_K^{-1} [\hat{h}_K, \hat{\bar{V}}^{\frac{1}{2}}] \right]. \nonumber \\
 &&
    \label{ClassicVol2Q}
      \end{eqnarray} 
At this point to obtain the spectrum of the inverse volume operator 
we express (\ref{ClassicVol2Q}) using the $SU(2)$ identity in (\ref{holonomiyI2})
and we arrive to the following form 
 \begin{eqnarray}
&&\widehat{\frac{\mbox{sgn}(E)}{\sqrt{\bar{V}}}}  =  - \frac{512 \, i}{3 \, l_P^4 \mu_0^3}  \, 
 \epsilon_{i j k}  \sum_{I J K} \epsilon^{I J K} \, n^i_I n^j_J n^k_K \, \nonumber \\
 && \hspace{4cm}\left[\cos \left(\frac{A_I}{2}\right) \, \hat{\bar{V}}^{\frac{1}{2}} \, \sin \left(\frac{A_I}{2} \right)
 - \sin \left(\frac{A_I}{2}\right) \, \hat{\bar{V}}^{\frac{1}{2}} \, \cos \left(\frac{A_I}{2} \right)\right] \nonumber \\
 && \hspace{4cm} \left[\cos \left(\frac{A_J}{2}\right) \, \hat{\bar{V}}^{\frac{1}{2}} \, \sin \left(\frac{A_J}{2} \right) 
 - \sin \left(\frac{A_J}{2}\right) \, \hat{\bar{V}}^{\frac{1}{2}} \, \cos \left(\frac{A_J}{2} \right) \right] \nonumber\\
 &&\hspace{4cm}  \left[\cos \left(\frac{A_K}{2}\right) \, \hat{\bar{V}}^{\frac{1}{2}} \, \sin \left(\frac{A_K}{2} \right) 
 - \sin \left(\frac{A_K}{2}\right) \, \hat{\bar{V}}^{\frac{1}{2}} \, \cos \left(\frac{A_K}{2} \right) \right]. \nonumber\\
 &&
  \label{ClassicVol2Q2} 
     \end{eqnarray} 
All the operators in brackets commute with the volume operator and so the operator
(\ref{ClassicVol2Q2}) is diagonal on the basis (\ref{basis}). Using the relation 
$\epsilon_{ijk}n^i_I n^j_J n^k_K = \epsilon_{IJK}$ it is straightforward to obtain the 
spectrum on the Hiilbert space basis     
\begin{eqnarray}
&& \widehat{\frac{\mbox{sgn}(E)}{\sqrt{\bar{V}}}} |\mu_E, \mu_{E^1} \rangle =
\frac{8}{\sqrt{2} l_P \mu_0^3} \,
|\mu_E|^{\frac{1}{2}} \, |\mu_{E^1}|^{\frac{1}{2}} \nonumber \\
&&\hspace{3cm}
 \left(|\mu_E + \mu_0|^{\frac{1}{4}} - 
|\mu_E - \mu_0|^{\frac{1}{4}}\right) \, \left(|\mu_{E^1} + \mu_0|^{\frac{1}{2}} - 
|\mu_{E^1} - \mu_0|^{\frac{1}{2}}\right)^2|\mu_E, \mu_{E^1} \rangle. \nonumber \\
&&
   \label{spectrum} 
     \end{eqnarray} 
The spectrum of the inverse volume operator is upper bounded near the classical
singularity which is in $E=0$ or $\mu_E=0$ and reproduce the correct classical 
spectrum of $1/\bar{V}$ for large volume eigenvalues ($\mu_E \sim \mu_{E^1} \gg 0$),
\begin{eqnarray}
 \widehat{\frac{\mbox{sgn}(E)}{\sqrt{\bar{V}}}} |\mu_E, \mu_{E^1} \rangle \rightarrow
\frac{2 \sqrt{2}}{l_P} \,
\frac{1}{|\mu_E|^{\frac{1}{4}} \, |\mu_{E^1}|^{\frac{1}{2}} }|\mu_E, \mu_{E^1} \rangle.
\label{limit}
\end{eqnarray}
We see that the spectrum of the inverse volume operator is discrete. It is not divergent
differing from the classical operator which has a singularity in $E=0$ 
and reproduce the correct classical behavior for large values of the volume eigenvalues.    

Now we pass to study the operator $1/\sqrt{|E|}$. The classical quantity defined 
in (\ref{CurvInv}) becomes the operator 
 \begin{eqnarray}
&&  \hspace{0cm}\widehat{\frac{\mbox{sgn}(E) \, \mbox{sgn}(E^1)}{\sqrt{|E|}} }= - \frac{64}{\mu_0^2 \, l_P^3}
 \, \mbox{Tr} \left[\tau_3 \hat{h}_1^{-1} [\hat{h}_1, \hat{\bar{V}}^{\frac{1}{2}}] \right] \, 
 \mbox{Tr} \left[(\tau_2+\tau_1) \hat{h}_2^{-1} [\hat{h}_2, \hat{\bar{V}}^{\frac{1}{2}}] \right] \nonumber \\
 && =  - \frac{64 \sqrt{2}}{\mu_0^2 \, l_P^3}
         \left[\cos\left(\frac{A \mu_0 l_P}{2}\right) \hat{\bar{V}}^{\frac{1}{2}} 
         \sin\left(\frac{A \mu_0 l_P}{2}\right) \hat{\bar{V}}^{\frac{1}{2}} - 
         \sin\left(\frac{A \mu_0 l_P}{2}\right) \hat{\bar{V}}^{\frac{1}{2}}
         \cos\left(\frac{A \mu_0 l_P}{2}\right) \hat{\bar{V}}^{\frac{1}{2}}\right] \nonumber \\
 && \hspace{0.97cm}\left[\cos\left(\frac{A_1 \mu_0 l_P}{\sqrt{2}}\right) \hat{\bar{V}}^{\frac{1}{2}} 
         \sin\left(\frac{A_1 \mu_0 l_P}{\sqrt{2}}\right) \hat{\bar{V}}^{\frac{1}{2}} - 
         \sin\left(\frac{A_1 \mu_0 l_P}{\sqrt{2}}\right) \hat{\bar{V}}^{\frac{1}{2}}
         \cos\left(\frac{A_1 \mu_0 l_P}{\sqrt{2}}\right) \hat{\bar{V}}^{\frac{1}{2}}\right].  \nonumber\\
         &&                  \label{CurvInvQuant}
 \end{eqnarray}
We consider the action of the operator (\ref{CurvInvQuant}) on the basis states of the Hilbert 
space and we obtain the spectrum of such operator  
\begin{eqnarray}
&& \hspace{-1cm}\widehat{\frac{\mbox{sgn}(E) \, \mbox{sgn}(E^1)}{\sqrt{|E|}} } \, |\mu_E, \mu_{E^1} \rangle =
\frac{\sqrt{2}}{\mu_0^2 l_P} \, |\mu_E|^{\frac{1}{4}} \, |\mu_{E^1}|^{\frac{1}{2}}\nonumber \\
&&\hspace{2cm}\left[ |\mu_{E^1} + \mu_0|^{\frac{1}{2}} - |\mu_{E^1} - \mu_0|^{\frac{1}{2}} \right]
\left[ |\mu_{E} + \mu_0|^{\frac{1}{4}} - |\mu_{E} - \mu_0|^{\frac{1}{4}} \right]
\, |\mu_E, \mu_{E^1} \rangle.
\label{spectrumInvCurv}
\end{eqnarray}
The spectrum of $1/\sqrt{|E|}$ is well defines in $E = 0$, where the
classical Schwarzschild singularity is localized. 

If we consider the limite of large eigenvalues 
$\mu_E \rightarrow \infty$ and $\mu_{E^1} \rightarrow \infty$ we obtain 
\begin{eqnarray}
&& \hspace{-1cm}\widehat{\frac{\mbox{sgn}(E) \, \mbox{sgn}(E^1)}{\sqrt{|E|}} } 
\, |\mu_E, \mu_{E^1} \rangle  \rightarrow \frac{\sqrt{2}}{l_P |\mu_E|^{\frac{1}{2}}}
\, |\mu_E, \mu_{E^1} \rangle.
\end{eqnarray}
This is the correct classical value of $1/\sqrt{|E|}$.

In this section we have studied the operators relevant for the singularity analysis 
in quantum gravity and we have found that the singularity disappears and that
the space-time can be extend beyond the classical singular point localized in $\mu_E = 0$.
 
 \subsection{Hamiltonian constraint and quantum dinamics}
In this section we study the dynamics of the model. We want resolve the Hamiltonian
constraint in the dual space of the Hilbert space.

The quantum version of the Hamiltonian constraint 
defined in (\ref{hamiltonianEreg2}) can be obtained promoting the classical holonomies
to operators and the poisson bracket to the commutator    
\begin{eqnarray}
 \hat{H}_E  =  \frac{8 \pi \, i}{\mu_0^3 \, \hbar} \, \sum_{I J K}\, \epsilon^{I J K} \, \mbox{Tr} \left[ \hat{h}_I \hat{h}_J \hat{h}_I^{-1} \hat{h}_J^{-1} \hat{h}_{[IJ]} \, \hat{h}_K^{-1} [\hat{h}_K, \hat{V} ]  \right].
\label{hamiltonianEreg2Q}
\end{eqnarray}
Using the relations in (\ref{holonomiyI2}) we can express the Hamiltonian operator as
\begin{eqnarray}
&& \hat{H}_E  =  - \frac{8 \pi \, i}{\mu_0^3 \, \hbar} \, 
 \Big[4 \sin(2x) \sin(2y) \left(\sin(y) \hat{V} \cos(y) - \cos(y) \hat{V} \sin(y) \right) +\nonumber \\
&& +2 \left( \cos(\beta) \sin^2(2y) - \sin(\beta) \left(\sin^2(2y) + \cos(2y)\right)\right)
\left(\sin(x) \hat{V} \cos(x) - \cos(x) \hat{V} \sin(x) \right) \Big],\nonumber \\
&&
 \label{hamiltonianEreg2Q2}
\end{eqnarray}
 where we have introduced the following notations: 
 $x = A \mu_0 l_P/2$, $y = \sqrt{2} A_1 \mu_0/2$ and $\beta = \mu_0^2/2$.
 
 Before calculating the solution of the Hamiltonian constraint we must consider
 the action of the operator $\sin(\dots)$ and $\cos(\dots)$ on the Hilbert space 
 for all possible trigonometric arguments which appear in the Hamiltonian
 constraint (\ref{hamiltonianEreg2Q2}),
 \begin{eqnarray}
&& \cos(x)|\mu_E, \mu_{E^1} \rangle = 
\frac{1}{2} \Big(|\mu_E + \mu_0, \mu_{E^1}  \rangle + |\mu_E - \mu_0, \mu_{E^1} \rangle \Big)\nonumber \\
&& \sin(x)|\mu_E, \mu_{E^1} \rangle = 
 \frac{1}{2i} \Big(|\mu_E + \mu_0, \mu_{E^1} \rangle- |\mu_E - \mu_0, \mu_{E^1} \rangle \Big)\nonumber \\
 && \cos(y)|\mu_E, \mu_{E^1} \rangle = 
\frac{1}{2} \Big(|\mu_E , \mu_{E^1}  + \mu_0 \rangle + |\mu_E, \mu_{E^1} - \mu_0\rangle \Big)\nonumber \\
&& \sin(y)|\mu_E, \mu_{E^1} \rangle = 
 \frac{1}{2i} \Big(|\mu_E, \mu_{E^1} + \mu_0 \rangle- |\mu_E, \mu_{E^1} - \mu_0\rangle \Big)\nonumber \\
 && \sin(2x)|\mu_E, \mu_{E^1} \rangle = 
 \frac{1}{2i} \Big(|\mu_E + 2 \mu_0, \mu_{E^1} \rangle- |\mu_E - 2 \mu_0, \mu_{E^1} \rangle \Big)\nonumber \\
&& \sin(2y)|\mu_E, \mu_{E^1} \rangle = 
 \frac{1}{2i} \Big(|\mu_E, \mu_{E^1} + 2 \mu_0 \rangle- |\mu_E, \mu_{E^1} - 2\mu_0\rangle \Big)\nonumber \\
 && \cos(2y)|\mu_E, \mu_{E^1} \rangle = 
 \frac{1}{2} \Big(|\mu_E, \mu_{E^1} + 2 \mu_0 \rangle+|\mu_E, \mu_{E^1} - 2\mu_0\rangle \Big)\nonumber \\
 && \sin^2(2y)|\mu_E, \mu_{E^1} \rangle = 
- \frac{1}{4} \Big(|\mu_E, \mu_{E^1} + 4 \mu_0 \rangle - 2 |\mu_E, \mu_{E^1}\rangle + |\mu_E, \mu_{E^1} - 4\mu_0\rangle \Big).\nonumber \\
 \label{cossin}
\end{eqnarray}

Using all these ingredients we can calculate the action of the Hamiltonian constraint 
on the Hilbert space basis  
\begin{eqnarray}
&& \hspace{-0.0cm} \hat{H}_E |\mu_E, \mu_{E^1} \rangle = \frac{8 \pi^2 l_P^3}{\sqrt{2} \, \mu_0^3 \hbar}
\Big[ 
- \alpha(\mu_E, \mu_{E^1})
\Big(|\mu_E+2\mu_0, \mu_{E^1}+2\mu_0 \rangle
- |\mu_E-2\mu_0, \mu_{E^1}+2\mu_0 \rangle \nonumber \\
&&\hspace{5.7cm} - |\mu_E+2\mu_0, \mu_{E^1}-2\mu_0 \rangle
+|\mu_E-2\mu_0, \mu_{E^1}-2\mu_0 \rangle\Big) \nonumber \\
&&\hspace{-0.09cm} + \beta(\mu_E, \mu_{E^1}) \Big(\frac{\sin(\mu_0^2/2) - \cos(\mu_0^2/2)}{2}\Big)
\Big(|\mu_E, \mu_{E^1}+4\mu_0 \rangle 
-2 |\mu_E, \mu_{E^1}\rangle +
|\mu_E, \mu_{E^1}-4\mu_0 \rangle \Big)  \nonumber \\
&& \hspace{4.3cm}-\beta(\mu_E, \mu_{E^1}) \sin(\mu_0^2/2)
\Big(|\mu_E, \mu_{E^1}+2\mu_0 \rangle 
+|\mu_E, \mu_{E^1}-2\mu_0 \rangle \Big)  \nonumber \\
&&
\end{eqnarray}

At this point we can resolve the Hamiltonian constraint. 
The solutions of the Hamiltonian constraint are
in the $\mathcal{C}^{\star}$ space that is the dual of the dense subspace $\mathcal{C}$ of the kinematical space. A generic element of this space is 
\begin{eqnarray}
\langle \psi | = \sum_{\mu_E, \mu_{E^1}} \psi(\mu_E, \mu_{E^1}) \langle \mu_E, \mu_{E^1}|.
\end{eqnarray}
The constraint equation $\hat{H} |\psi \rangle = 0$ is now interpreted as an equation in the dual space $\langle \psi | \hat{H}^{\dag}$;
from this equation we can derive a relation for the coefficients $\psi(\mu_E, \mu_{E^1})$  
\begin{eqnarray}
&& \hspace{-0.7cm}- \alpha(\mu_E - 2 \mu_0, \mu_{E^1} - 2 \mu_0) \, \psi(\mu_E - 2 \mu_0, \mu_{E^1} - 2 \mu_0) 
+ \alpha(\mu_E + 2 \mu_0, \mu_{E^1} - 2 \mu_0) \, \psi(\mu_E + 2 \mu_0, \mu_{E^1} - 2 \mu_0) \nonumber \\
&&\hspace{-0.7cm} + \alpha(\mu_E - 2 \mu_0, \mu_{E^1} + 2 \mu_0) \, \psi(\mu_E - 2 \mu_0, \mu_{E^1} + 2 \mu_0) 
- \alpha(\mu_E + 2 \mu_0, \mu_{E^1} + 2 \mu_0) \, \psi(\mu_E + 2 \mu_0, \mu_{E^1} + 2 \mu_0) \nonumber \\
&& \hspace{-0.5cm}+ \Big(\frac{\sin(\mu_0^2/2) -  \cos(\mu_0^2/2)}{2}\Big)
\Big(\beta(\mu_E, \mu_{E^1} - 4 \mu_0) \, \psi(\mu_E, \mu_{E^1} - 4 \mu_0) -
\beta(\mu_E, \mu_{E^1}) \, \psi(\mu_E, \mu_{E^1}) \nonumber \\
&& \hspace{3.5cm}+ \beta(\mu_E, \mu_{E^1} + 4 \mu_0) \, \psi(\mu_E, \mu_{E^1} + 4 \mu_0)\Big) \nonumber \\
&& \hspace{-0.5cm} - \sin(\mu_0^2/2) \Big(\beta(\mu_E, \mu_{E^1} - 2 \mu_0) \, \psi(\mu_E, \mu_{E^1} - 2 \mu_0) +
\beta(\mu_E, \mu_{E^1} +2 \mu_0) \, \psi(\mu_E, \mu_{E^1} +2 \mu_0)\Big) = 0,
\label{solution}
\end{eqnarray}
where the functions $\alpha(\mu_E, \mu_{E^1})$ and  $\alpha(\mu_E, \mu_{E^1})$
are define by 
\begin{eqnarray}
&& \alpha(\mu_E, \mu_{E^1}) \equiv |\mu_E|^{\frac{1}{2}}
\left(|\mu_{E^1} + \mu_0| - |\mu_{E^1} -\mu_0| \right) \nonumber \\
&& \beta(\mu_E, \mu_{E^1}) \equiv |\mu_{E^1}|
\left(|\mu_{E} + \mu_0|^{\frac{1}{2}} - |\mu_{E} -\mu_0|^{\frac{1}{2}} \right)
\label{alfabeta}
\end{eqnarray}
How can be seen from equation (\ref{solution}) the quantization program 
produce a difference equation and imposing a boundary condition
we can obtain the wave function $\psi(\mu_E, \mu_{E^1})$ for the black hole.
We can interpret $\psi(\mu_E, \mu_{E^1})$ as the wave function of the anisotropy 
 ``$E^1$" at the time  ``$E$". As in loop quantum cosmology also in this case the state $\psi(0,0)$ decouples from the dynamics.

\section*{Conclusions}

In this paper we studied the Kantowski-Sachs space time in Ashtekar variables.
We started from the complete ``loop quantum gravity" and we reduced  to the space-time
of spatial topology  $\mathbf{R} \times \mathbf{S}^2$. 
This space-time is very important
if we want to understand what happens very closed to the black hole singularity where the
quantum gravity effects are dominant and the classical Schwarzschild solution 
it isn't correct. 
In particular we concentrated on the Schwarzschild black hole solution 
inside the horizon where the time exchanges with space and the model 
reduces to a cosmological type space-time:
a general space-time with this topology is called a Kantowski-Sachs space-time.
 
 The quantization procedure is induced by the full ``loop quantum gravity".
We have introduced holonomies for the homogeneous space-time and we have expressed the 
inverse volume operator, the classically singular quantity $1/\sqrt{|E|}$ and the Euclidean 
part of the Hamiltoninan constraint in terms of holonomies. 
All these operators are well defined on the kinematical  Hilbert space. 
 
The main results are :

\begin{enumerate}
\item  the inverse volume operator has a finite spectrum in all the region
inside the horizon and we can conclude that the classical singularity disappears 
in  ``loop quantum gravity"; for large eigenvalues of the volume operator on the
other side we find the classical inverse volume behaviour,
\item the quantity $1/\sqrt{|E|}$, connected to the Schwarzschild 
curvature invariant $R_{\mu\nu \rho \sigma} R^{\mu\nu \rho \sigma} \sim 1/b(t)^6$
by  $b(t)^2 = |E|$,
is singularity free in quantum gravity and reproduces the classical behaviour 
for large eigenvalues,
\item  the solution of the Hamiltonian constraint gives a discrete difference
equation for the coefficients of the physical states which are defined in the 
dual space of the dense subspace of the kinematical Hilbert space.
\end{enumerate}
An important consequence of the quantization is that, unlike the classical 
evolution, the quantum evolution doesn't stop at the classical singularity
and the ``other side" of the singularity corresponds to a new domain where
the triad reverses its orientation. 
This work is useful if we want understand what is the mechanism 
to resolve the problem of the ``information loss" in the process of black 
hole formation \cite{AB}.

\section*{Acknowledgements}
We are grateful to Carlo Rovelli, Eugenio Bianchi and Guido Cossu for many important and clarifying discussion. This work is supported in part by a grant from the Fondazione Angelo Della Riccia.

\section*{Appendix}
We report some technical properties extensively used in this paper
and the independent components of the curvature $F_{ab}$ for the Kantowski-Sachs space time.

The $SU(2)$ generators used in the paper are $\tau^i = - \frac{i}{2} \sigma^i$, where $\sigma^i$ are the Pauli matrices 

\begin{equation}
\sigma^1= \left(\begin{array}{cc}
                          0     &     1    \\
                         1    &     0    \\

\end{array}   \right), \hspace{1cm}
\sigma^2= \left(\begin{array}{cc}
                          0     &     -i     \\
                         i       &     0    \\

\end{array}   \right), \hspace{1cm}
\sigma^3= \left(\begin{array}{cc}
                          1     &     0     \\
                          0     &     -1    \\
                          
\end{array}   \right).
\label{pauli}
\end{equation}

Other important properties of the Pauli matrix and of the $SU(2)$ generators $\tau^i$ are 
\begin{eqnarray}
&& [\sigma^i, \sigma^j] = 2 \,  i \epsilon^{ijk} \, \sigma^k, \nonumber \\
&& \sigma^i \, \sigma^j = \delta^{ij} + i \, \epsilon^{ijk} \, \sigma^k, \nonumber \\
&& [\tau^i, \tau^j] = 2 \,  i \epsilon^{ijk} \, \tau^k, \nonumber \\
&& \mbox{Tr}[\tau^i \tau^j] = - \frac{1}{2} \delta^{ij}, \nonumber \\
&& \mbox{Tr}[\tau^i \tau^j \tau^k] = - \frac{1}{4} \epsilon^{ijk}. 
\end{eqnarray}

The field strenght $F_{ab}$ independent components are 
\begin{eqnarray}
&& F_{12} = A A_1 \, (\tau_2 - \tau_1), \nonumber \\
&& F_{13} = -A A_1 \, (\tau_2 + \tau_1) \sin \theta, \nonumber \\
&& F_{23} = (2A_1^2 - 1) \, \tau_3 \sin \theta.
\end{eqnarray}

\end{document}